\makeatletter
\def\@ddllist#1#2{\ifx#1\empty \else
 \ifx#2\empty \let#2#1\else \edef#2{#1,#2}\fi \fi}
\def\@ddrlist#1#2{\ifx#1\empty \else
 \ifx#2\empty \let#2#1\else \edef#2{#2,#1}\fi \fi}
\let\addlefttolist=\@ddllist
\let\addrighttolist=\@ddrlist
\def\@reorder#1{\def\@newlst{}\edef\@tmplst{#1}\def\@qm{?}%
 \@for\@nxtdum:=\@tmplst\do{\def\@bigbit{}%
 \@for\@nxtchr:=\@tmplst\do{\ifx\@bigbit\empty\def\@tmplst{}\fi
 \ifx\@nxtchr\@qm \@ddrlist\@nxtchr\@newlst \else \ifx\@bigbit\empty
 \let\@bigbit=\@nxtchr \else \ifnum\@nxtchr>\@bigbit
 \@ddrlist\@bigbit\@tmplst \let\@bigbit=\@nxtchr
 \else \@ddrlist\@nxtchr\@tmplst \fi \fi \fi}%
 \@ddllist\@bigbit\@newlst}\edef#1{\@newlst}}
\def\@collapse#1{\def\@newlst{}\def\@qm{?}\def\@tmpbit{}\def\@link{,}%
 \@for\@nxtchr:=#1\do{\ifx\@newlst\empty \let\@newlst=\@nxtchr \else
 \ifx\@nxtchr\@qm
 \edef\@newlst{\@newlst\@tmpbit,\@nxtchr}\def\@tmpbit{}\def\@link{,}\else
 \ifx\@lstchr\@qm \edef\@newlst{\@newlst,\@nxtchr}\else
 \ifnum\@tempcnta<\@nxtchr
 \edef\@newlst{\@newlst\@tmpbit,\@nxtchr}\def\@tmpbit{}\def\@link{,}\else
 \ifnum\@tempcnta=\@nxtchr
 \edef\@tmpbit{\@link\@nxtchr}\def\@link{--}\fi\fi\fi\fi\fi
 \ifx\@nxtchr\@qm \else \@tempcnta=\@nxtchr \advance\@tempcnta\@ne \fi
 \let\@lstchr=\@nxtchr}\edef\@newlst{\@newlst\@tmpbit}\edef#1{\@newlst}}
\let\collapselist=\@collapse \let\reorderlist=\@reorder
\def\citesort#1{\edef\@lst{#1}\@reorder{\@lst}\@collapse{\@lst}\edef#1{\@lst}}
\def\@citex[#1]#2{\if@filesw\immediate\write\@auxout{\string\citation{#2}}\fi
 \def\@citea{}\@for\@citeb:=#2\do
 {\ifx\@citea\empty\else \edef\@citea{\@citea ,}\fi
 \@ifundefined{b@\@citeb}{\edef\@citea{\@citea?}%
 \@warning{Citation `\@citeb' on page \thepage\space undefined}}%
 {\edef\@citea{\@citea\csname b@\@citeb\endcsname}}}%
 \@reorder\@citea \@collapse\@citea \@cite{\@citea}{#1}}
\makeatother
\newcommand{\LQCD}{\Lambda_{\mbox{\tiny QCD}}}
\newcommand{\LQCDfour}{\LQCD^{\mbox{\tiny(4)}}}
\newcommand{\AQO}[1]{\bar#1\gamma_3\gamma_5#1}
\newcommand{\PME}[1]{\langle\,p^\uparrow\,|\;#1\;|\,p^\uparrow\,\rangle}
\newcommand{\AV}{{\cal A}}
\newcommand{\M}{\hphantom{-}} \newcommand{\Z}{\hphantom{0}}
\newcommand{\eg}{{\em e.g.}}  \newcommand{\ie}{{\em i.e.}}
\def\abbreviation#1#2#3{\def#1{#3 (#2)\def#1{#2}}}
\abbreviation {\BSR}   {BSR}  {Bjorken sum rule}
\abbreviation {\DIS}   {DIS}  {deep-inelastic scattering}
\documentstyle{qcdparis}
\begin{document}
\pagestyle{plain}

\title{Proton Spin, Sum Rules, QCD and Higher Twist
$^*$							
}
\author{Philip G. Ratcliffe}
\affil{Dip.\ di Fisica, Univ.\ di Milano\\
       via G. Celoria 16, 20133 Milano, Italy}

\abstract{We examine the present status of the Bjorken sum rule in the light of
recent data on the spin structure functions of the proton, neutron and deuteron
obtained by the CERN and SLAC experimental groups. We also discuss the r\^ole
of possible higher-twist contributions and higher-order PQCD corrections and
comment on the extraction of the necessary parameters, $D$ and $F$, obtained
from hyperon semi-leptonic decays.}

\resume{Nous examinons l'\'etat actuel de la r\`egle de somme de Bjorken \`a la
lumi\`ere des donn\'ees r\'ecentes sur les fonctions de structures du spin du
proton, neutron et deut\'eron obtenues par des groups exp\'erimentaux au CERN
et a SLAC. Nous discutons aussi le r\^ole des contributions \'eventuelles de
twist plus \'elev\'e et des corrections de PQCD a ordre plus \'elev\'e et nous
commentons l'extraction des param\`etres n\'ecessaires, $D$ et $F$, obtenus a
partir des d\'esint\'egrations semi-leptoniques des hyp\'erons.
\\[12pt]
\hfill hep-ph/9506383					
}

\twocolumn[%
\kern-1.5cm						
\maketitle]
\fnm{7} 						
{Talk given in the Polarisation Session at the Workshop
on Deep Inelastic Scattering and QCD (Paris, April 1995).}

\section{Introduction}
Polarization effects provide valuable insight into the dynamics of hadronic
interactions and can be sensitive to bound-state and other non-perturbative
physics. In particular, the \BSR~\cite{Bj} is a measurable quantity that may be
used to test theoretical predictions based on the light-cone expansion in PQCD.
The experimental precision now attainable is at the ten-percent level or better
while, on the theoretical side, all relevant PQCD calculations have been
carried out to at least two-loop order~\cite{Larin94} (\ie, approximately
one-percent level) and for the \BSR\ itself to three loops~\cite{Larin91}.

In the quark-parton model the structure function $g_1(x,Q^2)$~\cite{Review} is
related to polarized quark distributions, in a manner analogous to
$F_1(x,Q^2)$:
\begin{eqnarray}
 g_1(x,Q^2) &=& \mbox{$\frac{1}{2}$} \sum_f e_f^2 \, \Delta q_f(x,Q^2), \\
 F_1(x,Q^2) &=& \mbox{$\frac{1}{2}$} \sum_f e_f^2 \,        q_f(x,Q^2).
\end{eqnarray}
The polarized and unpolarized quark densities are defined in the following
manner:
\begin{eqnarray}
 \Delta q_f(x,Q^2) &=& q_f^+(x,Q^2) - q_f^-(x,Q^2), \\
        q_f(x,Q^2) &=& q_f^+(x,Q^2) + q_f^-(x,Q^2).
\end{eqnarray}
where $q_f^\pm(x,Q^2)$ are the densities of quarks of flavour $f$ and positive
or negative helicity with respect to the parent hadron.

Experimentally an asymmetry is measured and the polarized structure function is
then extracted via
\begin{equation}
 g_1(x,Q^2) = \frac {A_1(x,Q^2) \, F_2(x,Q^2)} {2x\,(1+R(x,Q^2))} ,
\end{equation}
where $R_1(x,Q^2)$ is the ratio of longitudinal to transverse unpolarized
structure functions and $A_1(x,Q^2)$ is the measured asymmetry.

\section{The Bjorken system of equations}
The full SU(3) algebra of the baryon octet admits three independent quantities,
which may be expressed in terms of the SU(3) axial-vector currents:
\begin{equation}
\begin{array}{r@{\;=\;}l}
 \AV_3 & \AQO{u} - \AQO{d}            , \\
 \AV_8 & \AQO{u} + \AQO{d} - 2\AQO{s} , \\
 \AV_0 & \AQO{u} + \AQO{d} + \AQO{s}  .
\end{array}
\end{equation}
And thus
\begin{equation}
 \PME{\AV_i} = g_i \qquad (i=3,8,0), \\
\label{BJsystem}
\end{equation}
The r.h.s.\ of (\ref{BJsystem}) for \mbox{$i=3,8$} corresponds to axial-vector
couplings accessible in hyperon semi-leptonic decays
(\mbox{$g_3=1.2573\pm0.0028$}~\cite{PDG94} and
\mbox{$g_8=0.629\pm0.039$}~\cite{PGR90}), but $g_0$, corresponding to the
flavour-singlet axial-vector current, is unknown. Thus, a prediction for the
proton integral alone is impossible. A further independent combination of $u$,
$d$ and $s$ axial-current matrix elements is accessible via $\nu$-$p$ elastic
scattering~\cite{Ahre87} and this would allow a prediction for single nucleon
targets. However, the precision of such data is still very poor.

The Bjorken sum rule~\cite{Bj}, with PQCD radiative corrections, then reads
\begin{eqnarray}
 \Gamma_1^{p-n} &=& \int_0^1\!dx\;g_1^{p-n}(x,Q^2) \\
                &=& \mbox{$\frac{1}{6}$}\,g_3\,
                    \left[1-\alpha_s/\pi-c_2(\alpha_s/\pi)^2-\dots\right],
\end{eqnarray}
where the Wilson coefficients, $c_n$, are known up to \mbox{$n=3$}. Note that
in what follows the higher-order corrections will be suppressed for simplicity,
but it should always be borne in mind that all expressions receive QCD
corrections known to at least second order.

\section{The \mbox{\boldmath$D$} and \mbox{\boldmath$F$} parameters}
Central then to the theoretical analysis are the values of the SU(3)
constants, $D$ and $F$, parametrizing the axial couplings, $g_i$, involved in
hyperon semi-leptonic decays~\cite{PGR90}. There has, over the past few years,
been considerable discussion on the validity of the standard approach to their
extraction from data~\cite{Donog93,HSD,Ehrns95,Licht95}. The problem is
essentially that of how to account reliably for SU(3) breaking.

A systematic approach to the problem was presented in
ref.~\cite{Donog93}, where the so-called ``recoil'' correction was taken into
account, together with possible differences in the strange and $u$, $d$
sea-quark wave-functions. Later, also taking advantage of the more precise data
available, it was shown that the latter correction is strongly disfavoured by
the data and that the former alone provides a very satisfactory account of
SU(3) breaking~\cite{PGR90}. The principle results of this analysis have since
been confirmed by various authors~\cite{HSD}.

Recently two new approaches have been proposed: the first is based on a
phenomenological parametrization of SU(3) breaking in the ratio $F/D$ on which
we shall comment shortly~\cite{Ehrns95}, the second has been discussed at this
meeting and the reader is referred to the paper appearing in these
proceedings~\cite{Licht95}.

In~\cite{Ehrns95} it was noticed that $F/D$, as inferred from the three axial
coupling constants extracted from angular/spin correlations alone, apparently
obeys a simple linear law in an SU(3)-breaking mass parameter defined by
\begin{equation}
\delta = \frac{m_i+m_f-m_p-m_n}{m_i+m_f+m_p+m_n},
\end{equation}
where $m_{i,f}$ are the initial- and final-state baryon masses and $m_{p,n}$
the reference proton and neutron masses.

Solely on this basis the authors propose extrapolating their fit to
\mbox{$\delta=0$}, for which they obtain
\begin{equation}
F/D = 0.40 \pm 0.07,
\end{equation}
where the rather large error directly reflects the extrapolation procedure
adopted. The interest in such a value is that it would allow the Ellis-Jaffe
sum rule to be saturated without recourse to strange-quark polarization. Note,
however, that it could not, of course, simultaneously explain any discrepancy
with the Bjorken (or neutron) sum rule.

There are two strong criticisms to be levelled at this approach: one is of a
theoretical nature and the other more experimental. The theoretical difficulty
has to do with the fundamental nature of the $D$ and $F$ parameters themselves;
these are universal constants describing the symmetric and antisymmetric parts
of the SU(3) couplings and appear with different Clebsch-Gordan coefficients in
the various decay matrix elements. The hidden implication of such an approach
is that, due to SU(3) breaking, $D$ and $F$ are miraculously renormalized in
just such a way as to preserve the particular combination, \mbox{$F+D=g_3$}.
Note that precisely the same combination also governs
\mbox{$\Xi\to\Sigma^0e\nu$}, which is thus predicted to remain {\em
un\/}renormalized despite the large value of the breaking parameter, $\delta$,
for this decay.

The other objection is against an arbitrarily selective use of data; the
decay-rate data are completely ignored and these both increase the overall
precision and shift the final answers, while also highlighting the inability of
this approach to globally describe the data well. Note that the decay-rate data
are both more numerous and more varied in their $D$-$F$ dependence than the
angular/spin correlation data alone. Let me then stress that a simple --
physically motivated -- recoil correction provides good agreement between all
present data and that the low value of $F/D$ proposed in ref.~\cite{Ehrns95}
would appear highly improbable~\cite{PGR95b}.

\section{The Ellis-Jaffe Sum Rule}
Arguments may be made for setting the strange-quark axial matrix element to
zero~\cite{Ellis74}: the strange quarks in the proton are very few and are
concentrated below \mbox{$x_B\simeq0.1$}, where all correlations with the
parent nucleon are expected to be very weak. Thus, the matrix elements in
eq.~(\ref{BJsystem}) for \mbox{$i=8$} and 0 should be approximately equal,
leaving only two independent quantities and allowing predictions for the proton
and neutron separately:
\begin{equation}
 \Gamma_1^{p(n)} = (-)\mbox{$\frac{1}{12}$} g_3
                    + \mbox{$\frac{5}{36}$} g_8
                    + \mbox{$\frac{1}{3}$}  \PME{\AQO{s}} ,
\end{equation}
where the last term is then assumed negligible. Conversely, given the value of
$\Gamma_1$, the value of either the strange-quark or singlet axial-vector
matrix elements may be extracted from these equations. There is no space here
to discuss in detail the strange-quark spin problem; the interested reader is
referred to~\cite{Prep88,Prep90a}, where a bound on the non-diffractive
component and thus on the strange-quark polarization was derived, references to
critiques of these papers may be found therein. In short, this analysis leads
to the following bound: \mbox{$\left|\Delta{s}\right|\le0.02$}. In what
follows, we shall not in fact place great emphasis on the bound, but recall it
here lest this physically intuitive result be forgotten.

\section{A Comparison of Theory and Experiment}
We now compare the experimental results with theoretical predictions based on
the framework outlined above. In performing the calculations we have used the
very precise value of $\LQCDfour$ recently extracted in a three-loop analysis
of scaling violations in \DIS~\cite{Pare94}, which is thus most suited to our
purposes. Such an analysis also allows an examination of the possible
improvement to be attained on increasing the order of the perturbative
corrections included. It should always be stressed that, for consistency, all
quantities must be evaluated at the same loop order and that, in particular, it
is meaningless to insert a two-loop $\alpha_s$ into a three-loop expression and
{\em vice versa\/}.
\begin{equation}
\begin{array}{lr@{\;=\;}l}
\mbox{EMC~\cite{EMC88}}  & \Gamma_1^p(11)
 & \M 0.128 \pm 0.010 \pm 0.015 \nonumber \\
\mbox{SMC~\cite{SMC94}}  & \Gamma_1^p(10)
 & \M 0.136 \pm 0.011 \pm 0.011 \nonumber \\
\mbox{E143~\cite{E143}}  & \Gamma_1^p(3)\Z
 & \M 0.127 \pm 0.004 \pm 0.010           \\
\mbox{SMC~\cite{SMC93}}  & \Gamma_1^d(5)\Z
 & \M 0.023 \pm 0.020 \pm 0.015 \nonumber \\
\mbox{E143~\cite{E143d}} & \Gamma_1^d(3)\Z
 & \M 0.049 \pm 0.004 \pm 0.003           \\
\mbox{E142~\cite{E142}}  & \Gamma_1^n(2)\Z
 &  - 0.022 \pm 0.006 \pm 0.009 \nonumber \\
[6pt]
 & \Gamma_1^p(11)
 & \M 0.182 \pm 0.006 + \mbox{$\frac13$}\Delta{s} \nonumber \\
 & \Gamma_1^p(10)
 & \M 0.182 \pm 0.006 + \mbox{$\frac13$}\Delta{s} \nonumber \\
\makebox[0cm][l]{\raisebox{ 6pt}[0pt][0pt]{Ellis-}}%
\makebox[0cm][l]{\raisebox{-6pt}[0pt][0pt]{Jaffe }}
 & \Gamma_1^p(3)\Z
 & \M 0.179 \pm 0.006 + \mbox{$\frac13$}\Delta{s}           \\
 & \Gamma_1^d(5)\Z
 & \M 0.085 \pm 0.006 + \mbox{$\frac13$}\Delta{s} \nonumber \\
 & \Gamma_1^n(2)\Z
 &  - 0.010 \pm 0.006 + \mbox{$\frac13$}\Delta{s}\nonumber  ,
\end{array}
\end{equation}
where the number in parenthesis refers to the mean value of $Q^2$ in GeV$^2$
and, where necessary, nuclear corrections have already been
included~\cite{SMC93,E143d}. The short-fall in the proton integral with
respect to the Ellis-Jaffe prediction (taking \mbox{$\Delta{s}=0$}) is evident.
This observation led to coining the phrase {\em Spin Crisis\/}. A similar
observation may be made for the E143 deuteron integral. In contrast, the
neutron sum rule appears well satisfied by the E142 data. Thus, in terms of the
strange-quark contribution, both the EMC and SMC data imply
\mbox{$\Delta{s}\simeq-0.15$} while that of E142 leads to
\mbox{$\Delta{s}\simeq-0.04$}.

A measure of the discrepancy between the data and theory may be obtained by
extracting, experiment-by-experiment, the singlet axial-vector matrix elements:
the results are
\begin{equation}
\begin{array}{r@{\;=\;}l@{\qquad}l}
\Delta\Sigma
 & 0.17 \pm 0.17 & \mbox{EMC  proton}     \\
 & 0.25 \pm 0.15 & \mbox{SMC  proton}     \\
 & 0.24 \pm 0.09 & \mbox{E143 proton}     \\
 & 0.09 \pm 0.25 & \mbox{SMC  deuteron}   \\
 & 0.32 \pm 0.05 & \mbox{E143 deuteron}   \\[6pt]
 & 0.46 \pm 0.10 & \mbox{E142 neutron}    \\
 & 0.31 \pm 0.05 & \mbox{global deuteron} \\
 & 0.23 \pm 0.07 & \mbox{global proton}   ,
\end{array}
\label{DSigma}
\end{equation}
where $\Delta\Sigma$ is the invariant sum of quark polarizations as in
eq.~(\ref{BJsystem}) with \mbox{$i=0$}, \ie, evaluated for \mbox{$Q^2=\infty$}.
Comparison of the last three lines of (\ref{DSigma}) reveals the nature of the
problem: unless a very large PQCD (or otherwise) correction is invoked the
proton data imply a significantly smaller value of $\Delta{q}$ than do those
for the neutron. We remark in passing that the SLAC deuteron data is perfectly
in line with the mean of the proton and neutron data; thus, providing
reassurance as to the validity of the theoretical nuclear corrections
introduced and attesting the overall consistency of the experimental picture.

Alternatively, the strange-quark spin contribution may be fit~\cite{Prep93b};
taking the SLAC proton and neutron data and performing completely consistent
fits at one- two- and three-loop order, we obtain respectively
\mbox{$\chi^2=3.7$}, 3.8 and 3.2 for one degree of freedom. Using the Particle
Data Group~\cite{PDG94} preferred value of \mbox{$\LQCDfour=260^{+56}_{-46}$}
in a two-loop fit (for consistency with the extraction of $\LQCD$), the
situation is marginally improved to give \mbox{$\chi^2=2.8$}.

\section{Higher-Twist Contributions}
Given the low $Q^2$ of the SLAC data, it is natural to worry about the
possibility of higher-twist ``contamination.'' Two approaches to this problem
are either to theoretically estimate the size of such effects (\eg, using a
bag model~\cite{Ji93} or QCD sum rules~\cite{Balit90,Ross94}) or to deduce
limits from the well-documented higher-twist behaviour of unpolarized
data~\cite{PGR93}. In both cases the magnitude of higher-twist contributions
found is too small to have any real impact, even on the SLAC neutron data (by
an odd quirk, the higher-twist contribution to $g_1^n$ is typically much
smaller even than that in the case of $g_1^p$).

Furthermore, note that while the inclusion of large higher-twist contributions
can in principle restore agreement between predictions and data as far as the
sum-rule integrals at fixed $Q^2$ are concerned, the data seem to prefer only
mild (logarithmic) scaling violations. A similar situation has already been
noted in the case of the Gross-Llewellyn Smith sum rule~\cite{Gross69}, where
the size of corrections (perturbative or not) required by the sum rule is
larger than, and incompatible with, that deduced from the $Q^2$
variation~\cite{Kataev94}.

We should also mention an approach based on the known limiting behaviour for
\mbox{$Q^2\to0$}~\cite{Ansel89,Burke92}. The resulting effects are found to be
rather large and even in conflict with the observed $Q^2$ dependence. Moreover,
this analysis depends crucially on an assumed smooth interpolation through the
low-$Q^2$ resonance region.

\section{A Possible Explanation and Consequences}
It is interesting to ask what happens if the normalization condition on the
Wilson coefficients is relaxed, \ie, if PQCD is ignored and current algebra is
used only to fix ratios of matrix elements~\cite{PGR93}. In this case, adopting
our strange-quark bound to effectively set \mbox{$\Delta{s}=0$}, any one data
set may be used to fix the overall normalization. The SLAC proton data, for
example, then lead to the following ``prediction'' for the neutron:
\begin{equation}
0.002 \le \Gamma_1^n \le -0.026,
\end{equation}
in rather good agreement with the SLAC neutron data.

Alternatively, the quark spins may be deduced from the proton and neutron data
(with absolutely {\em no} assumption on the strange-quark spin) and the
following relation is then obtained:
\begin{equation}
\Gamma_1^n = -\mbox{$\frac1{11}$}\Gamma_1^p + \mbox{$\frac23$}\Delta s,
\end{equation}
which leads to a strange-quark polarization of
\begin{equation}
\Delta{s}=-0.03\pm0.03,
\end{equation}
a non-trivial result perfectly compatible with the bound. The total spin of the
quarks is then found to be
\begin{equation}
\Delta\Sigma=0.33\pm0.03.
\end{equation}
Thus, a satisfactory and self-consistent picture may be rendered, provided we
are willing to admit non-perturbative effects in the overall normalization of
the relevant operator matrix elements. That such contributions might be
important should not be too surprising given that the ``real'' intermediate
quark and gluon states, on which the PQCD Wilson-coefficient calculations are
based, clearly do not correspond to the physical hadronic states of the real
world.

\section*{Aknowledgements}
It is a pleasure to thank the convenors of the polarization working group and
the organizing committee.


\end{document}